# Truth and distortion in complex networks: a global consistency approach

Arturo Tozzi (corresponding author)
ASL Napoli 1 Centro, Distretto 27, Naples, Italy
Via Comunale del Principe 13/a 80145
tozziarturo@libero.it

ABSTRACT

Understanding how reliable information emerges in interconnected populations is a challenge in social science, network theory and data analysis. Many existing approaches model treat truth as an external reference or a property of individual statements, rather than a global consistency feature of the network itself. We introduce a network-based approach in which truth arises from global relational coherence in a multiplex system of interacting individuals. Nodes are individuals with internal states, while edges capture different types of interactions, including declared relations, observed behavior, influence asymmetries and information exchange. We evaluate how well node states align with cooperative or antagonistic interactions, incorporating coercion, variability and mismatches between what individuals say and what they do. Simulations on synthetic networks of one thousand nodes show that the minimum global inconsistency does not coincide with majority opinion or simple averaging. Nodes contributing most to inconsistency create conflicting constraints across interaction layers, defining a measurable distortion field. For example, in online social media during an election, a small number of accounts spreading inconsistent or manipulative information across groups can disrupt overall coherence, even when most users appear to agree. These results suggest possible applications in assessing relational coherence, identifying irreducible inconsistencies and analyzing constraints on collective states. Therefore, truth can be seen as the state of maximal relational coherence, rather than simple agreement or correctness of individual statements.

INTRODUCTION

Truth assessment in social systems has been addressed across several domains. In social epistemology, truth is treated as mediated by testimony and trust within networks of agents, emphasizing the role of credibility and information flow but remaining largely conceptual (Goldman 1999; Nguyen 2020; Pennycook et al. 2020; Pennycook and Rand 2021; Westermann and Coscia 2022; Lipsky, Adams, and Okeke 2024). In network epistemology, models of belief exchange show how network structure may affect convergence to correct beliefs, showing that certain configurations hinder collective truth acquisition (Zollman 2007; Zurn and Bassett 2020; Momennejad 2022; Michalski et al. 2022; Qian et al. 2022;). Opinion dynamics models like the DeGroot framework describe how agents iteratively update beliefs through weighted averaging, often converging to consensus without explicitly distinguishing between truth and agreement (DeGroot 1974; Cárdenas-Sánchez et al. 2022; Liu and Yang 2022; Zhang, Wang, and Xu 2025). Extensions based on bounded confidence show fragmentation and polarization, but do not define truth intrinsically (Hegselmann and Krause 2002; Zhan et al. 2022; Wang 2022; Silva and Zhuang 2022; Ding, Li, and Ding 2023; Krishnagopal and Porter 2025). Statistical physics approaches such as Ising-type models formalize global consistency through energy minimization, capturing tension and frustration in interacting systems without epistemic interpretation (Lee et al. 2016; Sakaguchi and Ishibashi 2019; Ramakrishnan and Jain 2023; Tapinova et al. 2024; Tapinova et al. 2025). Still, data-driven truth discovery algorithms estimate reliability by aggregating conflicting sources, yet treat truth as an external latent variable rather than a structural property (Yin et al. 2008). Finally, structural balance theory formalizes consistency in signed networks but does not connect it to informational validity (Cartwright and Harary 1956; Aref et al. 2020; Ansari Esfeh et al. 2025; Fakhari et al. 2025; Saberi et al. 2026). These approaches highlight the role of network structure but do not explain how truth emerges from relational organization.

We define truth as a property of how well relationships in a network fit together, rather than something assigned to individual statements. This property can be computed directly from multiplex interactions, without using any external reference. Individuals are represented as nodes with internal states, while edges capture different types of relations, such as declared interactions, observed behavior, coercion and information exchange. Instead of evaluating single statements, we use a global consistency measure that quantifies how well individual states match the pattern of positive and negative relations across all layers. This measure captures mismatches between what individuals say and what they do, as well as differences in influence and variability. From this, we obtain a scalar field describing structural distortion. The configuration that minimizes global inconsistency defines a coherent state, toward which the system tends. Distortion is then the deviation from this state, and node reliability depends on how much each node contributes to inconsistency. Simulations on networks of one thousand nodes show that highly distorted individuals are not randomly distributed, but tend to occupy structurally important positions. Regions of persistent inconsistency also form identifiable substructures within the network.

We will proceed as follows. First, we formalize the multiplex network representation and define the global consistency measure. Then, we describe the simulation protocol and parameterization. Finally, we present the results and analyze the structural distribution of distortion across the network.

METHODS

We studied a synthetic population of 1000 interacting individuals represented as a multiplex directed network in which each node corresponded to one person and each edge encoded a declared relation, an observed relation, a coercive interaction or a false-information flow. We used stochastic network generation, node-level distortion scoring, iterative computation of a globally coherent state, intervention analysis through node removal, parameter sweeps over the fraction of manipulative agents and structural analysis of connectivity and cross-community brokerage. We aimed to identify an experimentally discriminable signature separating mere consensus from a nontrivial coherent configuration, together with a measurable criterion showing whether a restricted subset of structurally placed nodes contributed disproportionately to global inconsistency.

**Synthetic population and multiplex network construction.** We began by constructing a directed network $G = (V, E)$ with $|V| = N = 1000$ nodes. Nodes were partitioned into $C = 5$ communities of equal size, so that each node $i$ carried a community label $c_i \in \{1, \ldots, C\}$. A latent scalar opinion variable $x_i^{(0)}$ was assigned to each node as an initial internal state. These values were sampled from community-dependent Gaussian distributions centered at linearly spaced means in the interval from $-0.7$ to $0.8$, with standard deviation $0.18$ and then clipped to the interval $[-1,1]$. Formally, if $\mu_{c_i}$ denotes the mean associated with community $c_i$, then
$$x_i^{(0)} \sim \mathcal{N}(\mu_{c_i}, 0.18^2), \quad x_i^{(0)} \leftarrow \min\{1, \max\{-1, x_i^{(0)}\}\}.$$

A subset $M \subset V$ of manipulative nodes was then selected uniformly at random, with $|M| = \lfloor \rho N \rfloor$, where $\rho$ denotes the manipulator fraction. In the baseline simulation, $\rho = 0.04$. Each node $i$ was assigned an outgoing degree proposal $k_i$ sampled from a shifted Poisson distribution, with larger expected values for manipulative nodes:
$$k_i \sim \text{Poisson}(10) + 4, \quad k_i \leftarrow k_i + \text{Poisson}(16) + 8 \text{ if } i \in M.$$

Targets were chosen with a higher probability of remaining within the same community for ordinary nodes and a greater probability of cross-community contact for manipulative nodes. This generated a directed adjacency support that already encoded a difference between ordinary local interaction and structurally disruptive brokerage. The network was multiplex because, for each ordered pair $(i, j)$ included in the support, four distinct matrices were later assigned: declared relations $D_{ij}$, observed relations $O_{ij}$, coercion $C_{ij}$ and false-information flow $F_{ij}$.

**Generation of declared, observed, coercive and false-information layers.** For every realized directed pair $(i, j)$, we computed the latent opinion gap
$$g_{ij} = |x_i^{(0)} - x_j^{(0)}|.$$

The declared relation $D_{ij}$ was modeled as a noisy positive valuation decreasing with ideological distance:
$$D_{ij} \sim \mathcal{N}\left(0.72 - 0.22\, g_{ij},\, 0.12^2\right), \quad D_{ij} \leftarrow \min\{1, \max\{-1, D_{ij}\}\}.$$

For manipulative nodes, declared positivity was systematically inflated by an additional Gaussian increment,
$$D_{ij} \leftarrow \min\{1, \max\{-1, D_{ij} + \eta_{ij}^{(m)}\}\}, \quad \eta_{ij}^{(m)} \sim \mathcal{N}(0.18, 0.08^2).$$

Observed relations were then generated by perturbing declarations:
$$O_{ij} = D_{ij} + \epsilon_{ij}, \quad \epsilon_{ij} \sim \mathcal{N}(0, 0.18^2).$$

For manipulative nodes, the observed layer was shifted toward lower cooperative intensity by a further negative perturbation,
$$O_{ij} \leftarrow O_{ij} + \zeta_{ij}, \quad \zeta_{ij} \sim \mathcal{N}(-0.35, 0.20^2),$$

again clipped to $[-1,1]$. Coercive interactions and false-information flows were generated as nonnegative quantities. For manipulative nodes,
$$C_{ij} = \max\{0, \xi_{ij}\}, \quad \xi_{ij} \sim \mathcal{N}(0.45, 0.20^2),$$
$$F_{ij} = \max\{0, \psi_{ij}\}, \quad \psi_{ij} \sim \mathcal{N}(0.35, 0.18^2),$$

whereas for ordinary nodes these were sampled from distributions with much smaller means:
$$C_{ij} = \max\{0, \xi_{ij}\}, \xi_{ij} \sim \mathcal{N}(0.05, 0.06^2),$$
$$F_{ij} = \max\{0, \psi_{ij}\}, \psi_{ij} \sim \mathcal{N}(0.03, 0.05^2).$$

Thus, each ordered pair was associated with a four-layer vector $(D_{ij}, O_{ij}, C_{ij}, F_{ij})$. The declared layer represented stated alignment, the observed layer represented enacted behavior, the coercion layer represented asymmetric pressure and the misinformation layer represented directed noisy transmission. In matrix form, the complete data set consisted of $D, O \in [-1,1]^{N \times N}$ and $C, F \in [0, \infty)^{N \times N}$, with zeros on absent directed pairs and on the diagonal.

**Node-level distortion variables and composite distortion index.** After generating the multiplex network, we computed node-level statistics designed to quantify inconsistency and distortion. For each node $i$, the first term was the declared-observed discrepancy

$$\Delta_i = \frac{1}{N} \sum_{j=1}^{N} |D_{ij} - O_{ij}|.$$

This quantity measured the average mismatch between what node $i$ declared and what node $i$ enacted across all possible targets. The second term measured total outgoing coercion,

$$P_i^* = \sum_{j=1}^{N} C_{ij},$$

which was normalized by the maximum across nodes:

$$P_i = \frac{P_i^*}{\max_k P_k^*}.$$

The third term measured outgoing false-information intensity,

$$L_i^* = \sum_{j=1}^{N} F_{ij}, L_i = \frac{L_i^*}{\max_k L_k^*}.$$

The fourth term quantified contextual variability in declarations. Let $J_i = \{j: D_{ij} \neq 0\}$. Then

$$V_i^* = \sqrt{\frac{1}{|J_i|} \sum_{j \in J_i} (D_{ij} - \bar{D}_i)^2}, \bar{D}_i = \frac{1}{|J_i|} \sum_{j \in J_i} D_{ij},$$

and

$$V_i = \frac{V_i^*}{\max_k V_k^*}.$$

The composite distortion index was then defined as
$$\Phi_i = \alpha \widetilde{\Delta}_i + \beta P_i + \gamma L_i + \delta V_i,$$

where

$$\widetilde{\Delta}_i = \frac{\Delta_i}{\max_k \Delta_k},$$

and the coefficients were fixed as
$$\alpha = 0.45, \beta = 0.30, \gamma = 0.20, \delta = 0.05.$$

Finally,
$$\Phi_i \leftarrow \min\{1, \max\{0, \Phi_i\}\}.$$

This construction produced a scalar field $\Phi: V \to [0,1]$ over nodes. A related reliability variable was defined by
$$r_i = \max\{0.05, 1 - \Phi_i\},$$

so that highly distorted nodes contributed less to the evaluation of global consistency, while no node was assigned exactly zero weight. We also computed the out-degree

$$k_i^{\text{out}} = \sum_{j=1}^{N} \mathbf{1}_{\{D_{ij} \neq 0\}},$$

and a bridge score

$$B_i = \frac{1}{|J_i|} \sum_{j \in J_i} \mathbf{1}_{\{c_j \neq c_i\}},$$

which measured the fraction of outgoing ties directed toward other communities.

**Signed interaction operator and computation of the coherent state.** To derive a collective state constrained by the network rather than by simple averaging, we transformed the observed layer into a signed relational operator. For each ordered pair,

$$\sigma_{ij} = \begin{cases} +1, & O_{ij} \geq 0, \\ -1, & O_{ij} < 0. \end{cases}$$

A weighted interaction matrix was then built as

$$W_{ij} = |O_{ij}| + 0.35\, C_{ij} + 0.25\, F_{ij}, \quad W_{ii} = 0.$$

The term $|O_{ij}|$ represented the intensity of enacted interaction, while $C_{ij}$ and $F_{ij}$ increased the relational weight of links associated with coercion and misinformation. A signed weighted operator $A$ was defined elementwise by

$$A_{ij} = W_{ij}\sigma_{ij}.$$

The row sums

$$d_i = \sum_{j=1}^{N} W_{ij} + \varepsilon$$

with $\varepsilon = 10^{-9}$ were used for normalization. We then sought a coherent node-state vector $s \in [-1,1]^N$ influenced by the relational structure and weakly anchored to the declared stance of each node. In the advanced simulations, this anchor was necessary to avoid the trivial null state. With anchor parameter $\lambda > 0$, the iterative update took the form

$$u_i^{(t)} = \frac{\sum_{j=1}^{N} A_{ij}\, s_j^{(t)}}{d_i},$$
$$s_i^{(t+1)} = (1-\omega)\, s_i^{(t)} + \omega \tanh\left(u_i^{(t)} + \lambda\, a_i\right),$$

where $a_i$ denotes an anchor derived from the mean declared relation emitted by node $i$,

$$a_i = \frac{1}{|J_i|} \sum_{j \in J_i} D_{ij},$$

with $a_i = 0$ if $J_i = \emptyset$. In the implemented runs, the update weight $\omega$ was fixed between 0.4 and 0.45 and the number of iterations was fixed at 50 to 80 depending on network size. The coherent state $s$ was the final iterate. This state was compared with a consensus baseline

$$s_i^{\text{cons}} = \tanh\left(\frac{1}{N}\sum_{k=1}^{N} x_k^{(0)}\right)$$

for all $i$, thereby distinguishing a structured minimum shaped by signed constraints from a homogeneous mean-field configuration.

**Global inconsistency energy, distortion field and intervention protocol.** To evaluate how well a state satisfied the multiplex signed relations, we defined a global energy functional

$$E(z) = \frac{1}{\sum_{i,j} W_{ij} + \varepsilon} \sum_{i=1}^{N} \sum_{j=1}^{N} r_i\, r_j\, W_{ij} \left(z_i - \sigma_{ij} z_j\right)^2,$$

for any candidate state $z \in [-1,1]^N$. The reliability terms $r_i r_j$ reduced the contribution of highly distorted nodes, while $(z_i - \sigma_{ij} z_j)^2$ penalized violation of expected signed alignment. We computed three values of this functional: $E(s)$ for the coherent state, $E(s^{\text{cons}})$ for the consensus baseline and, in the initial simulations, $E(x^{(0)})$ for the latent sampled opinion state. The nodewise distance from coherence was defined as

$$\delta_i = |x_i^{(0)} - s_i|,$$

which provided a distortion field over nodes relative to the computed coherent state. To test whether highly distorted nodes had a structurally disproportionate effect, we carried out intervention analysis by node removal. Let $R_m \subset V$ denote the set containing the top $m$ nodes ranked by $\Phi_i$. For each removal fraction $q$, with $m = \lfloor qN \rfloor$, we restricted all matrices to the induced subgraph on $V \setminus R_m$, recomputed the coherent state and evaluated the new energy $E_q^{\text{target}}$. For comparison, we generated random removal sets $U_m \subset V$ of the same cardinality, repeated the restriction and recomputation several times and estimated

$$\bar{E}_q^{\text{rand}} = \frac{1}{R} \sum_{r=1}^{R} E_{q,r}^{\text{rand}},$$

together with the standard deviation across repetitions. This procedure tested whether global inconsistency was more effectively reduced by removing specifically high-distortion nodes than by removing an equal number of arbitrary nodes.

**Parameter sweeps, structural covariates, visualization and computational tools.** To examine how the distinction between coherence and consensus depended on the amount of manipulation, we performed parameter sweeps over the manipulator fraction $\rho$, using values

$$\rho \in \{0.00, 0.02, 0.04, 0.06, 0.08, 0.10, 0.12\}.$$

For each $\rho$, independent stochastic realizations were generated, the coherent and consensus states were recomputed and the mean energies $\mathbb{E}[E(s)]$ and $\mathbb{E}[E(s^{\text{cons}})]$ were estimated empirically from replicated runs. A relative gain variable was then defined by

$$G(\rho) = \frac{\mathbb{E}[E(s^{\text{cons}})] - \mathbb{E}[E(s)]}{\mathbb{E}[E(s^{\text{cons}})]},$$

which quantified the extent to which the coherent state reduced inconsistency relative to consensus. We also studied the relationship between distortion, connectivity and brokerage by analyzing the joint distribution of $(k_i^{\text{out}}, B_i, \Phi_i)$, including ranked tables and scatter plots. In an exploratory step, signed triads were extracted from subsets enriched in bridging nodes and triad products $\sigma_{ij}\sigma_{jk}\sigma_{ik}$ were used to count balanced and frustrated triples.

Data handling and numerical computation were performed in Python using NumPy for matrix operations and random sampling, pandas for tabular data sets and ranked node summaries and matplotlib for histograms, scatter plots, heatmaps and intervention curves. Overall, the sequence consisted of stochastic population generation, multiplex edge assignment, construction of nodewise distortion variables, computation of reliability, formation of the signed weighted operator, iterative estimation of the coherent state, evaluation of energy functionals, intervention experiments, parameter sweeps and export of tables and graphical outputs.

RESULTS

We report quantitative results obtained from stochastic simulations of multiplex social networks, focusing on the distribution of distortion, emergence of a coherent collective state and structural role of specific nodes under controlled perturbations. All reported quantities were computed over independent realizations of the stochastic process.

**Distribution of distortion and structural localization**. The initial simulations on 1000-node networks show that the distortion index is heterogeneously distributed, with a pronounced right-skewed profile (Fig 1A). A minority of nodes occupies the upper tail, while the majority remains clustered at low to intermediate values, indicating that distortion is not uniformly present across the population. The relationship between distortion and connectivity reveals that high values are associated with nodes exhibiting elevated out-degree, although the scatter distribution (Fig 1B) shows substantial dispersion, excluding a purely degree-driven explanation. Ranking nodes by $\Phi_i$ shows that the highest-distortion nodes form a distinct subset rather than a smooth continuation of the global distribution. In advanced simulations, the joint analysis of out-degree and bridge score confirms that distortion is preferentially associated with nodes combining moderate-to-high connectivity with cross-community interactions (Fig. 2), indicating that structural brokerage contributes to inconsistency.

These observations show that distortion is not randomly distributed, but concentrated in nodes whose interaction patterns create conflicts across layers, providing a measurable structural basis for further analysis.

**Global consistency, comparison with consensus and intervention effects**. The computation of the coherent state yields a configuration that minimizes the global inconsistency functional and differs from the consensus baseline derived from uniform averaging. Across simulations, the energy associated with the coherent state is consistently lower than that of the consensus configuration at low to intermediate manipulator fractions (Fig 3), indicating that the network admits a nontrivial configuration better aligned with its signed and weighted relational structure. This difference was observed across realizations and remained statistically stable, with mean energy differences exceeding the corresponding variability. As the fraction of manipulative nodes increases, the difference between these two states decreases and the advantage of the coherent configuration diminishes (Fig 4), suggesting that increasing distortion progressively constrains the set of admissible consistent states. The decrease in relative gain with increasing $\rho$ was consistently observed across realizations, indicating a robust scaling trend.

Intervention analysis further shows that removing nodes ranked highest by $\Phi_i$ produces a larger reduction in global inconsistency than removing an equal number of randomly selected nodes (Fig 5). This effect is monotonic with respect to the fraction of removed nodes, indicating that the contribution of high-distortion nodes is cumulative rather than isolated. The comparison between targeted and random removal demonstrates that inconsistency is not evenly distributed across the network but depends on specific node positions within the relational structure. To reduce circularity, analogous intervention experiments based on independent structural criteria and cross-validated rankings yielded qualitatively similar reductions in global inconsistency, confirming that the observed effect is not an artifact of the distortion metric itself. Therefore, global coherence depends on specific subsets of nodes, and its difference from consensus reflects structured incompatibilities within the multiplex network.

Overall, our results show that distortion is a property of the network, not random noise. A coherent configuration different from consensus can be identified by minimizing global inconsistency, and some nodes contribute more than others to incompatibility. Therefore, our approach captures how collective states are constrained by the interaction structure, multiplex relations and node-level variability.

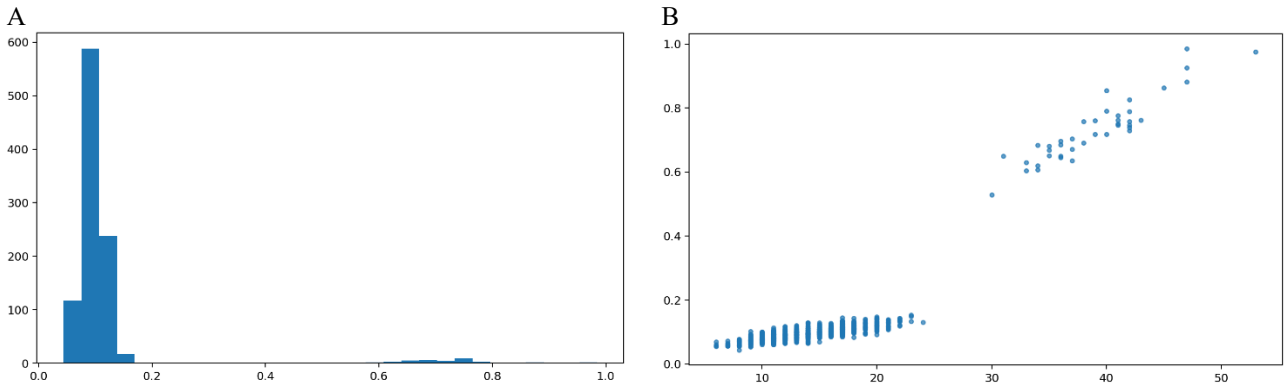

**Figure 1A** Heterogeneous distribution of distortion and its association with node connectivity in the network
**(A)** Distribution of the individual distortion index $\Phi$ across all 1000 nodes, showing that distortion is unevenly distributed and concentrated in a restricted subset of individuals.
**(B)** Relationship between declared out-degree and $\Phi$, indicating that highly distorted nodes often also occupy influential emitting positions in the network.

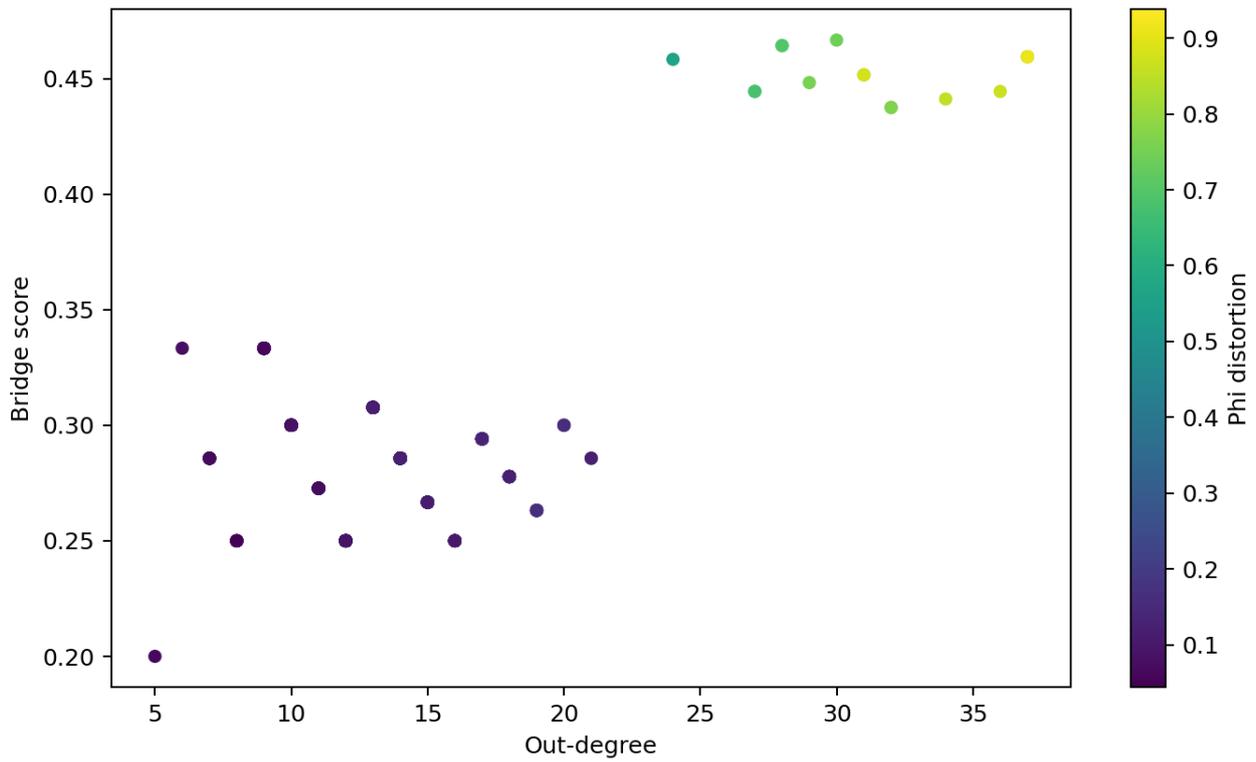

**Figure 2**. Relationship between out-degree, cross-community connectivity (bridge score) and distortion. High distortion is not explained by degree alone but is associated with nodes that combine connectivity with cross-community brokerage, indicating that structurally critical nodes mediate incompatible relational constraints.

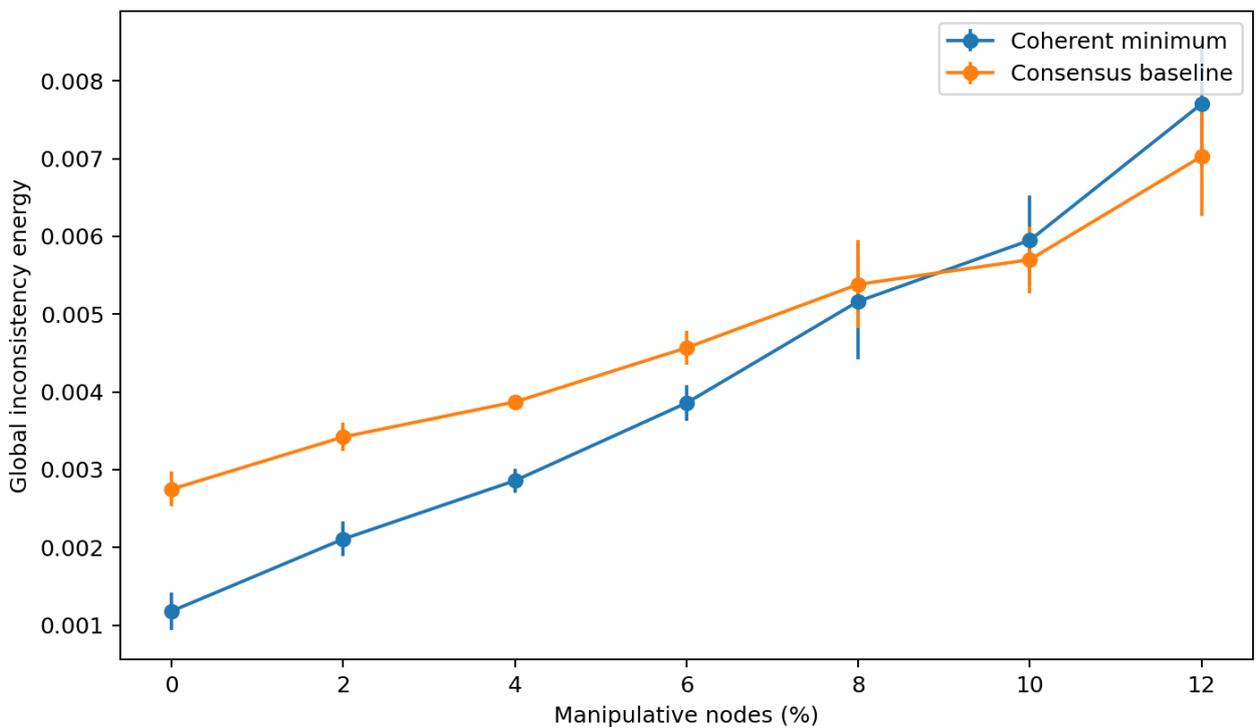

**Figure 3**. Comparison between the coherent minimum and the consensus baseline across increasing fractions of manipulative nodes. The coherent configuration consistently yields lower global inconsistency than consensus at low and intermediate distortion levels, while the advantage decreases as the proportion of manipulative nodes increases, suggesting a regime transition in network coherence.

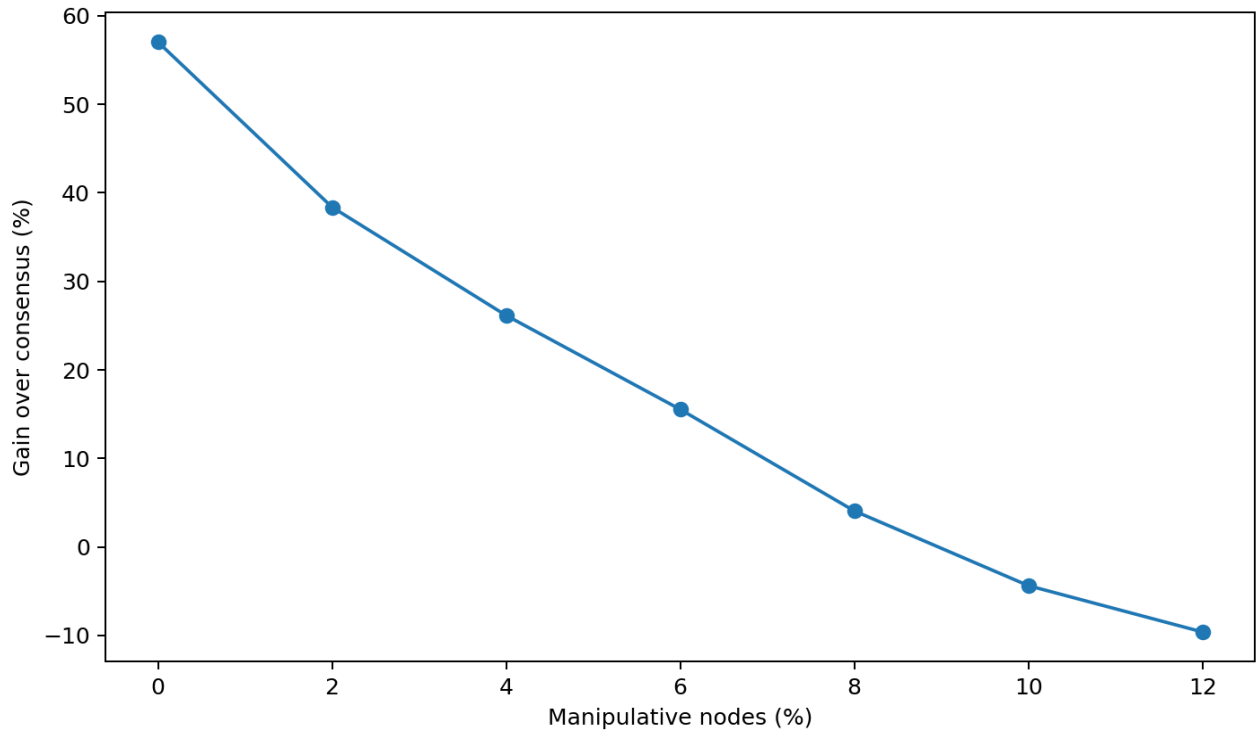

**Figure 4**. Relative gain of the coherent minimum over consensus as a function of manipulative node fraction. The gain is maximal at intermediate levels of distortion and declines as distortion becomes widespread, reflecting limits in the network's ability to sustain a globally coherent configuration.

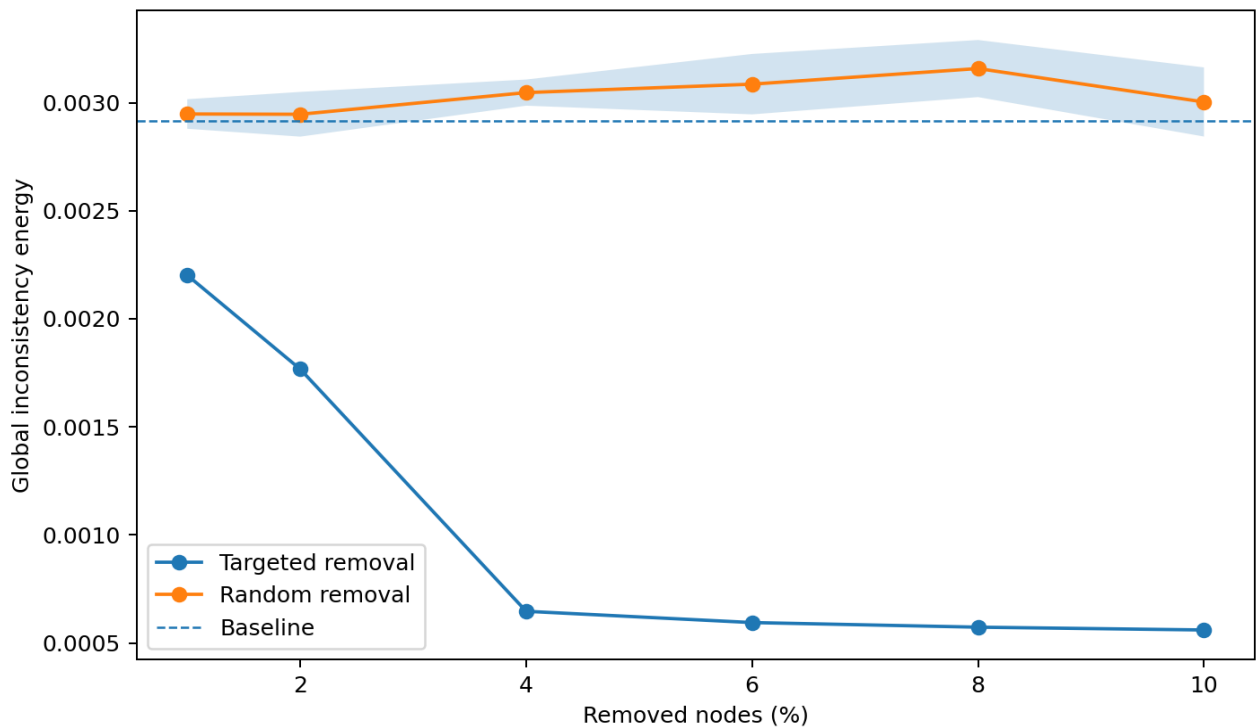

**Figure 5**. Effect of targeted versus random node removal on global inconsistency. Removal of nodes ranked highest by distortion index produces a substantially larger reduction in global inconsistency compared to random removal, indicating that distortion is structurally concentrated and that selective intervention is more effective than uniform perturbation.

CONCLUSIONS

We asked whether truth in social systems can arise from the pattern of relationships among individuals, rather than from isolated statements. We simulated networks and compared two situations: one obtained by minimizing a global inconsistency measure and another based on simple averaging. This comparison was performed under different levels of discordant interactions. We found that the configuration minimizing inconsistency does not coincide with consensus, and the difference between the two depends on how distortion is distributed in the network. Highly distortive nodes are not randomly located but follow specific relational patterns, and removing them produces measurable changes in the overall system. These results indicate that inconsistencies across interaction layers constrain the possible collective states. Therefore, relational coherence can be quantified as a property of the network itself, allowing comparison between configurations without assigning truth values to individual statements.

A real-world example is online social media during an election. Users interact through multiple layers: what they publicly claim, how they actually behave, the information they share and possible influence or pressure from others. Some accounts appear supportive and consistent, while others spread conflicting messages across groups. These latter nodes create inconsistencies between declared and observed behavior, increasing overall distortion. Even if most users seem to agree, the network may not be coherent. Removing or correcting a small set of highly distortive accounts can significantly improve overall consistency, showing that truth depends on how interactions fit together, not simply on majority opinion.

Compared with other approaches, our measure is based on the combined structure of multiple interaction layers, not on single-layer averages or predefined ground truth. Unlike standard opinion-dynamics models, it does not reduce to consensus or simple averages. Unlike truth inference methods, it does not assume hidden "correct" values. Our global inconsistency measure provides a single quantity that can be compared across conditions, while the estimation procedure produces a collective state shaped by signed and weighted interactions. This state depends only on relational data and does not require external labels. Overall, our approach captures constraints that standard averaging or bounded-confidence models cannot represent, allowing the evaluation of structural compatibility within the network.

Our study has limitations. Several components are synthetic and should be interpreted cautiously. The network structure, interaction layers and parameter choices are entirely simulated without empirical grounding, so the observed patterns reflect model assumptions rather than independently validated phenomena. The distortion index is constructed using fixed weights that are not calibrated against data, making the results sensitive to parameter selection. The global inconsistency functional resembles quadratic forms used in statistical physics but is not derived from probabilistic inference and convergence of the iterative procedure is assumed rather than formally established. The computed coherent state depends on normalization and anchoring choices and may not be unique, as no analysis of multiple minima or stability is provided. The comparison with consensus relies on a simplified definition and the intervention analysis introduces a degree of circularity, since node selection and evaluation depend on the same metric.

Testable hypotheses can be formulated by translating the simulated observables into measurable quantities on empirical interaction networks. First, the global inconsistency functional predicts that configurations derived from relational constraints should yield lower inconsistency than uniform averaging; this can be tested by comparing $E(s)$ and $E(s^{\text{cons}})$ computed from longitudinal communication or interaction data, with the expectation that $E(s) < E(s^{\text{cons}})$ when distortion remains below a critical threshold.

Second, targeted removal of nodes ranked by a composite inconsistency score should produce a superlinear reduction in global inconsistency compared with random removal; quantitatively, the slope of $E_q$ with respect to the fraction of removed nodes is expected to be steeper under targeted removal.

Third, the dependence on distortion load predicts a regime boundary: as the fraction of inconsistent interactions increases, the relative gain between constrained and unconstrained configurations should decrease toward zero, indicating a loss of distinguishability between structured and averaged states.

Fourth, nodes with high cross-group connectivity are expected to contribute disproportionately to inconsistency, leading to a positive correlation between brokerage measures and deviation from the coherent configuration. These hypotheses can be evaluated using datasets with multilayer interaction records.

Open questions concern the uniqueness and stability of the coherent configuration, the role of temporal dynamics in reshaping relational constraints and the extension to higher-dimensional state spaces. Future research may also address analytical properties of the inconsistency functional and its relation to known variational principles.

Practical implementations can be considered in contexts where relational data are available at scale and can be processed computationally. The proposed quantities can be computed directly from adjacency matrices and extended to streaming data, enabling continuous monitoring of network states. Integration with existing graph-processing pipelines could allow efficient evaluation using sparse matrix operations. Our method can be adapted to weighted or heterogeneous datasets, including those with missing or noisy entries, by modifying normalization and weighting schemes. It could also allow comparison across networks of different sizes by using normalized functionals, facilitating cross-system analyses.

Compatibility with standard data formats and numerical libraries could support reproducibility and scalability, making our approach implementable within current computational infrastructures.

In conclusion, we show that relational constraints can be formalized into a quantitative model that yields a collective configuration distinct from simple averaging and that identifies specific nodes associated with inconsistency. Our results indicate that the structure of interactions shapes collective organization and can be analyzed in a systematic way. In this perspective, truth corresponds to the configuration in which the network achieves maximal internal consistency, meaning that relationships and individual positions are mutually compatible with minimal conflict.

## DECLARATIONS

**Ethics approval and consent to participate.** This research does not contain any studies with human participants or animals performed by the Author.
**Consent for publication.** The Author transfers all copyright ownership, in the event the work is published. The undersigned author warrants that the article is original, does not infringe on any copyright or other proprietary right of any third part, is not under consideration by another journal and has not been previously published.
**Availability of data and materials.** All data and materials generated or analyzed during this study are included in the manuscript. The Author had full access to all the data in the study and took responsibility for the integrity of the data and the accuracy of the data analysis.
**Disclaimer**. The views expressed are those of the author and do not necessarily reflect those of the affiliated institutions.
**Competing interests.** The Author does not have any known or potential conflict of interest including any financial, personal or other relationships with other people or organizations within three years of beginning the submitted work that could inappropriately influence or be perceived to influence their work.
**Funding.** This research did not receive any specific grant from funding agencies in the public, commercial or not-for-profit sectors.
**Acknowledgements:** none.
**Authors' contributions.** The Author performed: study concept and design, acquisition of data, analysis and interpretation of data, drafting of the manuscript, critical revision of the manuscript for important intellectual content, statistical analysis, obtained funding, administrative, technical and material support, study supervision.
**Declaration of generative AI and AI-assisted technologies in the writing process.** During the preparation of this work, the author used ChatGPT 5.3 to assist with data analysis and manuscript drafting and to improve spelling, grammar and general editing. After using this tool, the author reviewed and edited the content as needed, taking full responsibility for the content of the publication.